\documentclass[preprint,showpacs,preprintnumbers,amsmath,amssymb,prb]{revtex4}
\usepackage{amsmath}
\usepackage{graphicx}
\usepackage{dcolumn}
\usepackage{bm}
\usepackage{mathrsfs}
\usepackage{hyperref}

\begin{document}

\title{Vortices in Quantum R\"ontgen Effect}

\author{Yi-Shi Duan}
\author{Ru-Nan Huang}
\thanks{Corresponding author}%
\email{huangrn04@lzu.cn}%
\affiliation{Institute of Theoretical Physics, Lanzhou University,
Lanzhou 730000, P. R. China}

\begin{abstract}

By the application of $\phi$-mapping topological theory, the
properties of vortices in quantum R\"ontgen effect is thoroughly
studied. The explicit expression of the vorticity is obtained,
wherein which the $\delta$ function indicates that the vortices can
only stem from the zero points of $\phi$ and the magnetic flux of
the consequent monopoles is quantized in terms of the Hopf indices
and Brouwer degrees. The evolution of vortex lines is discussed. The
reduced dynamic equation and a conserved dynamic quantity on stable
vortex lines are obtained.

\end{abstract}

\pacs{03.75.Lm, 03.70.+k, 47.32.-y \\
 Keywords: R\"ontgen effect; vortices; quantized monopoles}

\maketitle

\section{Introduction}\label{Sec1}

Intensive studies, both theoretical and experimental, have been
carried out since the advent of Bose-Einstein condensate \cite{1,2}.
Owing to its unique features, novel methods of investigation have
been springing up during the last decades. By virtue of recent
theory on quantum phase of induced dipoles \cite{3,4,5,6}, U.
Leonhardt and P. Piwnicki studied quantum R\"ontgen effect via a
mean-field approach (the Gross-Pitaevskii theory) and get some
intriguing results on its quantized monopoles \cite{7}.

Given a charged capacitor and a dielectric (such as glass, robber,
etc.) disc placed parallel between the plates, clearly the disc
would be polarized, and, when it rotates, the induced charges on its
opposite surfaces would act as currents and then generate a magnetic
field. This effect, which is known as R\"ontgen Effect, was
discovered by W.C. R\"ontgen in 1888 \cite{8}. Recent interests in
R\"ontgen effect were reborn for its promising perspective in the
research of quantum gases. For example, if the ordinary dielectric
disc is replaced by a quantum dielectric disc (a Bose-Einstein
condensate), Leonhardt and Piwnicki showed in their paper that only
vortices would generate a magnetic field and the field would behave
like it originates from a set of magnetic monopoles \cite{8}.
Despite their ingenuity, they employed the traditional way to
express the wave function
\begin{equation}\label{1}
 \psi=|\psi|e^{iS},
\end{equation}
which inevitably leads to the irrotational result
\begin{equation}
 \nabla\times\vec{u}=\nabla\times(\frac{\hbar}{m}\nabla S)=0,
\end{equation}
and causes an irreconcilable flaw to the whole theory.

In this paper, by the application of $\phi$-mapping topological
theory \cite{9}, which was established by Prof. Duan and has been
exercised profoundly in studying topological invariant and
topological structure of various physics systems
\cite{10,11,12,13,14}, we will not only solve the problem we
mentioned above but provide a thorough study on the behavior of
vortices in quantum R\"ontgen effect. The results we get could be
very conducive to further theoretical and experimental investigation
of BEC.

This paper is organized as follows: In Sec. \ref{Sec2}, we first
elucidate that the curl of the velocity field is
\begin{equation}
 \nabla\times\vec{u}=\frac{h}{m}\vec{J}
 (\frac{\phi}{x})\delta^2(\vec\phi),
\end{equation}
which indicates the magnetic monopoles are generated from the zero
points of $\psi$  (i.e. the locations of vortices). While those zero
points are regular points, the magnetic flux of the monopoles are
quantized in terms of the Hopf indices $\beta_i$  and Brouwer
degrees $\eta_i$ of $\phi$-mapping.

In Sec. \ref{Sec3} we study the critical points of $\psi$ , i.e.,
the limit points and bifurcation points, and show that there exists
branch processes. The vortex lines generate, annihilate, split or
merge at such points, while their topological number - winding
number $\beta\eta$ is conserved.

In Sec. \ref{Sec4} we investigate the reduced dynamic equation and
get a conserved dynamic quantity on stable vortex lines.

In Sec. \ref{Sec5} we draw our conclusions.

SI units are used throughout the paper.

\section{The Quantized Vorticity and Magnetic Flux}\label{Sec2}

Our work begins with the Gross-Pitaevskii equation which one could
readily obtain from the Lagrangian density of Ref. \cite{7}
\begin{equation}\label{4}
 i\hbar\frac{\partial\psi}{\partial t}
 =\frac{1}{2m}(\frac{\hbar}{i}\nabla+\alpha\vec{E}\times\vec{B})^2\psi
 -\frac{\alpha}{2}E^2\psi+\frac{g}{2}|\psi|^2\psi+V\psi.
\end{equation}
As declared by Leonhardt and Piwnicki, the constant $\alpha$ denotes
the electrical polarizability of the condensed atoms; the
Gross-Pitaevskii term $g|\psi|^2\psi$ $(g>0)$ models the atomic
repulsion (collision), which tends to smooth out density variations
over the healing length $\hbar/\sqrt{2gm|\psi|^2}$; and $V$ stands
for the external potential which prevents the condensate from
spreading out to infinity.

Instead of adopting the traditional expression (\ref{1}), we shall
write
\begin{equation}\label{5}
 \psi=\phi^1+i\phi^2,
\end{equation}
where
\begin{equation}
 \phi^a=\phi^a(x),\quad a=1,2,
\end{equation}
are real functions. It means $\psi$ can be described by a
2-dimensional vector field $\vec\phi=(\phi^1,\phi^2)$.

Substituting (\ref{5}) into (\ref{4}), we get the dynamic equation
of the polarized condensed atoms
\begin{eqnarray}\label{7}
 -\hbar\varepsilon_{ab}\phi^a\frac{\partial}{\partial t}\phi^b
 &=&-\frac{\hbar^2}{2m}\phi^a\nabla^2\phi^a
 +\frac{\hbar}{m}(\alpha\vec{E}\times\vec{B})\cdot\varepsilon_{ab}\phi^a\nabla\phi^b
 \nonumber\\
 &&+[\frac{1}{2m}(\alpha\vec{E}\times\vec{B})^2-\frac{\alpha}{2}E^2
 +\frac{g}{2}\|\phi\|^2+V]\|\phi\|^2,
\end{eqnarray}
as well as the continuity relation
\begin{equation}
 \frac{\partial}{\partial t}\|\phi\|^2+\nabla\cdot
 [(\frac{\hbar}{m}\frac{1}{\|\phi\|^2}\varepsilon_{ab}\phi^a\nabla\phi^b
 +\frac{\alpha}{m}\vec{E}\times\vec{B})\|\phi\|^2]=0,
\end{equation}
where
\begin{equation}
 \|\phi\|^2=\phi^a\phi^a=\psi^\ast\psi
\end{equation}
describes the density of the condensed atoms.

Consequently, the velocity field takes the form
\begin{equation}\label{10}
 \vec{u}
 =\frac{\hbar}{m}\frac{1}{\|\phi\|^2}\varepsilon_{ab}\phi^a\nabla\phi^b
 +\frac{\alpha}{m}\vec{E}\times\vec{B}.
\end{equation}
In the following study, the second term is omitted for its
contribution to the outcome is negligibly small (which was shown in
\cite{7}).

Note that
\begin{equation}
 \frac{\phi^a}{\|\phi\|^2}
 =\frac{1}{\|\phi\|}\frac{\partial}{\partial\phi^a}\|\phi\|
 =\frac{\partial}{\partial\phi^a}\ln\|\phi\|,
\end{equation}
curl $\vec{u}$ can be expressed as
\begin{equation}\label{12}
 \nabla\times\vec{u}
 =\frac{\hbar}{m}(\nabla\phi^1\times\nabla\phi^2)
 \frac{\partial^2}{\partial\phi^a\partial\phi^a}\ln\|\phi\|.
\end{equation}
By defining the Jacobian vector
\begin{equation}
 \vec{J}(\frac{\phi}{x})=\nabla\phi^1\times\nabla\phi^2,
\end{equation}
and utilizing the Laplacian relation in $\phi$ space \cite{15}
\begin{equation}
 \frac{\partial^2}{\partial\phi^a\partial\phi^a}\ln\|\phi\|
 =2\pi\delta^2(\vec\phi),
\end{equation}
(\ref{12}) can be reduced as
\begin{equation}
 \nabla\times\vec{u}
 =\frac{h}{m}\vec{J}(\frac{\phi}{x})\delta^2(\vec\phi).
\end{equation}
This explicit expression is the utter solution to the problem which
we mentioned in the introduction. Actually, such a $\delta$
-functional behavior of the vorticity has been assumed by Leonhardt
and Piwnicki, but they failed to give any proof.

From the relation between $\vec{H}$ and $\vec{u}$ \cite{7}
\begin{equation}
 \nabla\cdot\vec{H}
 =\varepsilon_0\chi\vec{E}\cdot(\nabla\times\vec{u}),
\end{equation}
we have
\begin{equation}
 \nabla\cdot\vec{H}
 =\frac{h}{m}\varepsilon_0\chi\vec{E}\cdot\vec{J}(\frac{\phi}{x})\delta^2(\vec\phi).
\end{equation}

Therefore, one can easily reach the conclusion that
$\nabla\cdot\vec{H}\neq0$ (i.e.$\nabla\cdot\vec{u}\neq0)$ if and
only if $\vec\phi=0$, the magnetic monopoles are generated from the
zero points of $\vec\phi$(i.e. the locations of vortices). This
makes the solution of $\vec\phi=0$ extremely significant.

Suppose the vector field $\vec\phi(x)$ has $N$ zero points
$\vec{z}_i(i=i,\ldots,N)$. The implicit function theorem assures us,
while $\vec{z}_i(i=i,\ldots,N)$ are the regular points of
$\vec\phi(x)$, i.e. $\vec{J}(\frac{\phi}{x})|_{\vec{z}_i}\neq0$,
they can be expressed as parameterized singular strings
\begin{equation}
 L_i:\quad\vec{z}_i=\vec{z}_i(t,s),\quad i=1,\ldots,N.
\end{equation}
These $N$ isolated singular strings are just the vortex lines.

For a fixed $t$, one can get the following result from
$\delta$-function theory \cite{15}
\begin{equation}
 \delta^2(\vec\phi)
 =\sum_{i=1}^{N}\beta_i
 \int_{L_i}\frac{\delta^3(\vec{x}-\vec{z}_i(s))}{|J(\phi/u)|_{\Sigma_i}}ds,
\end{equation}
where $\Sigma_i$ is the $i$th planer element transversal to $L_i$
with local coordinates $(u^1,u^2)$,
$J(\frac{\phi}{u})=\frac{\partial(\phi^1,\phi^2)}{\partial(u^1,u^2)}$,
and $\beta_i$ is a positive integer known as the Hopf index of
$\phi$-mapping.

Since the direction vector of $L_i$ is given by
\begin{equation}
 \frac{d\vec{x}}{ds}\bigg|_{\vec{z}_i}
 =\frac{\vec{J}(\phi/x)}{J(\phi/u)}\bigg|_{\vec{z}_i},
\end{equation}
we then have
\begin{equation}
 \nabla\times\vec{u}
 =\frac{h}{m}\sum_{i=1}^{N}\beta_i\eta_i
 \int_{L_i}\frac{d\vec{x}}{ds}\delta^3(\vec{x}-\vec{z}_i(s))ds,
\end{equation}
where $\eta_i=sgnJ(\frac{\phi}{u})=\pm1$ is the Brouwer degree, and
it characterizes the direction of the $i$th vortex line. This
expression shows the important inner topological structure of
$\nabla\times\vec{u}$.

Thus, the vorticity of the condensate is
\begin{equation}
 \Gamma=\int_\Sigma\nabla\times\vec{u}\cdot d\vec{S}
 =\frac{h}{m}\sum_{i=1}^{N}\beta_i\eta_i,
\end{equation}
and the magnetic flux of the consequent monopoles can be worked out
as
\begin{equation}
 \Phi=\int\mu_0\nabla\cdot\vec{H}dv
 =\frac{h}{mc^2}\chi U\sum_{i=1}^{N}\beta_i\eta_i
 =\frac{\chi}{c^2}U\Gamma.
\end{equation}
Here $U$ denotes the applied voltage of the capacitor, $\chi$ is the
susceptibility. It's obvious that both $\Gamma$ and $\Phi$ are
quantized by the product $\beta\eta$, which is called the winding
number.

\section{The Evolution of Vortex Lines}\label{Sec3}

The discussion we've been carrying on so far is under the condition
that $\vec{z}_i(i=1,\ldots,N)$ are regular points, but what happens
when some $\vec{z}^\ast=(t^\ast,\vec{x}^\ast)$ are critical points
(i.e. $\vec{J}(\frac{\phi}{x})\big|_{\vec{z}^\ast}=0$)? Soon we will
show, such vortex lines would no longer be stable - it would evolve.
In other words, generating and annihilating of vortex lines, as well
as splitting and merging, would take place.

If we still insist on using the implicit function theorem, we first
hope that one of the following Jacobians
\begin{equation}\label{24}
 D^i(\frac{\phi}{x})\bigg|_{\vec{z}^\ast}
 =\frac{\partial(\phi^1,\phi^2)}{\partial(t,x^i)}\bigg|_{\vec{z}^\ast},
 \quad i=1,2,3,
\end{equation}
is nonzero.

For simplicity, let's fix the $z$ coordinate and consider the locus
of $\vec{w}^\ast=(t^\ast,x^\ast,y^\ast)$.

\textbf{Case 1.} At least one of (\ref{24})'s Jacobians is nonzero.
(Such $\vec{z}^\ast$ is called the limit point of $\vec\phi(x)$.)
Suppose $D^2(\frac{\phi}{x})\big|_{\vec{z}^\ast}\neq0$, near
$\vec{w}^\ast$ we could locally solve $\vec\phi=0$ to get
\begin{equation}\label{25}
 t=t(x),\quad y=y(x).
\end{equation}
By differentiating the identity $\vec\phi(x,t(x),y(x))=0$ with
respect to $x$, we yield
\begin{equation}
 \frac{dt}{dx}=-\frac{J^3(\phi/x)}{D^2(\phi/x)},\quad
 \frac{dy}{dx}=-\frac{D^1(\phi/x)}{D^2(\phi/x)}.
\end{equation}
Then (\ref{25}) can be expanded at $\vec{w}^\ast$ as
\begin{equation}
 t=t^\ast+\frac{dt}{dx}\bigg|_{\vec{w}^\ast}(x-x^\ast)
 +\frac{1}{2}\frac{d^2t}{dx^2}\bigg|_{\vec{w}^\ast}(x-x^\ast)^2
 +o(|x-x^\ast|^2),
\end{equation}
i.e.,
\begin{equation}\label{28}
 t-t^\ast
 =\frac{1}{2}\frac{d^2t}{dx^2}\bigg|_{\vec{w}^\ast}(x-x^\ast)^2,
\end{equation}
which is a parabola on the $xt$ plane.

Eq.(\ref{28}) reveals that there exists branch processes at the
limit point. If $\frac{d^2t}{dx^2}\big|_{\vec{w}^\ast}>0$, we have
the branch solutions $x_1(t)$ and $x_2(t)$ for $t>t^\ast$, which
represent the generating process. If
$\frac{d^2t}{dx^2}\big|_{\vec{w}^\ast}<0$, we have the branch
solutions $x'_1(t)$ and $x'_2(t)$ for $t<t^\ast$, which represent
the annihilating process.

Besides, Eq.(\ref{28}) also tells us a simple approximate relation
near $\vec{w}^\ast$
\begin{equation}
 |x-x^\ast|\propto|t-t^\ast|^\frac{1}{2},
\end{equation}
from which we can get the generating/annihilating speed
\begin{equation}
 v\propto|t-t^\ast|^{-\frac{1}{2}}.
\end{equation}

No mater which process takes place, from the identity
$\nabla\cdot(\nabla\times\vec{u})=0$, one can always reach the
conclusion that the winding number $\beta\eta$ is conserved
\begin{equation}
 \beta_1\eta_1+\beta_2\eta_2=0,
\end{equation}
which indicates that the two vortex lines have the same Hopf index
and opposite directions.

\textbf{Case 2.} All of (\ref{24})'s Jacobians are zero. (Such
$\vec{z}^\ast$ is called the bifurcation point of $\vec\phi(x)$.)
Suppose one of the partial derivatives is nonzero. Let, for example,
$\frac{\partial\phi^1}{\partial y}\neq0$. Again, according to the
implicit function theorem, near $\vec{w}^\ast$ we could locally
solve $\phi^1=0$ to get
\begin{equation}
 y=y(t,x).
\end{equation}
Let $F(t,x)=\phi^2(t,x,y(t,x))$, from
\begin{equation}
 J^3(\frac{\phi}{x})\bigg|_{\vec{z}^\ast}=0,\quad
 D^2(\frac{\phi}{x})\bigg|_{\vec{z}^\ast}=0,
\end{equation}
one can respectively prove
\begin{equation}
 \frac{\partial F}{\partial x}\bigg|_{\vec{z}^\ast}=0,\quad
 \frac{\partial F}{\partial t}\bigg|_{\vec{z}^\ast}=0.
\end{equation}
Then the Taylor expansion of $F(x,t)$ in the neighborhood of
$\vec{w}^\ast$ can be expressed as
\begin{equation}
 F(x,t)=\frac{1}{2}A(x-x^\ast)^2+B(x-x^\ast)(t-t^\ast)
 +\frac{1}{2}C(t-t^\ast)^2+\cdots,
\end{equation}
where
\begin{equation}
 A=\frac{\partial^2F}{\partial x^2},\quad
 B=\frac{\partial^2F}{\partial x \partial t},\quad
 C=\frac{\partial^2F}{\partial t^2}.
\end{equation}
Note that
\begin{equation}
 F(t^\ast,x^\ast)=\phi^2(t^\ast,x^\ast,y(t^\ast,x^\ast)),
\end{equation}
we have
\begin{equation}
 A\left(\frac{dx}{dt}\bigg|_{\vec{w}^\ast}\right)^2
 +2B\frac{dx}{dt}\bigg|_{\vec{w}^\ast}+C=0
 \qquad \bigg(\textrm{or}\quad
 C\left(\frac{dt}{dx}\bigg|_{\vec{w}^\ast}\right)^2
 +2B\frac{dt}{dx}\bigg|_{\vec{w}^\ast}+A=0\bigg).
\end{equation}

1) If $A\neq0$, $B^2-4AC>0$, then
\begin{equation}
 \left(\frac{dx}{dt}\bigg|_{\vec{w}^\ast}\right)_1
 =\frac{-B+\sqrt{B^2-AC}}{A},\quad
 \left(\frac{dx}{dt}\bigg|_{\vec{w}^\ast}\right)_2
 =\frac{-B-\sqrt{B^2-AC}}{A},
\end{equation}
which simply represent the intersection of two vortex lines.

2) If $A\neq0$, $B^2-4AC=0$, then
\begin{equation}
 \left(\frac{dx}{dt}\bigg|_{\vec{w}^\ast}\right)_1
 =\left(\frac{dx}{dt}\bigg|_{\vec{w}^\ast}\right)_2
 =-\frac{B}{A},
\end{equation}
which could be either one vortex line splits into two vortex lines,
or two vortex lines merge into one vortex line, with the speed
$v=-\frac{B}{A}$. Same reason as before, the winding number
$\beta\eta$ is conserved in the process, i.e.
\begin{equation}
 \beta_1\eta_1+\beta_2\eta_2=\beta\eta.
\end{equation}

3) If $A=0$, $B\neq0$, then
\begin{equation}
 \left(\frac{dx}{dt}\bigg|_{\vec{w}^\ast}\right)=-\frac{C}{2B},
\end{equation}
which simply describes the motion of one vortex line.

\section{The Reduced Dynamic Equation and Conserved Dynamic Quantity on Stable Vortex Lines}\label{Sec4}

In Sec. \ref{Sec2} we have obtained the dynamic equation of the
polarized condensed atoms (\ref{7})
\begin{eqnarray}
 -\hbar\varepsilon_{ab}\phi^a\frac{\partial}{\partial t}\phi^b
 &=&-\frac{\hbar^2}{2m}\phi^a\nabla^2\phi^a
 +\frac{\hbar}{m}(\alpha\vec{E}\times\vec{B})\cdot\varepsilon_{ab}\phi^a\nabla\phi^b
 \nonumber\\
 &&+[\frac{1}{2m}(\alpha\vec{E}\times\vec{B})^2-\frac{\alpha}{2}E^2
 +\frac{g}{2}\|\phi\|^2+V]\|\phi\|^2.
\end{eqnarray}

Note that
\begin{equation}
 \frac{1}{\|\phi\|^2}\phi^a\nabla^2\phi^a
 =\frac{\nabla^2\|\phi\|}{\|\phi\|}
 -\left(\nabla\frac{\phi^a}{\|\phi\|}\right)^2,
\end{equation}
we then have
\begin{eqnarray}\label{45}
 \hbar\frac{1}{\|\phi\|^2}\varepsilon_{ab}\phi^a\frac{\partial}{\partial t}\phi^b
 &=&-\frac{\hbar^2}{2m}\left(\nabla\frac{\phi^a}{\|\phi\|}\right)^2
 -\frac{\hbar}{m}(\alpha\vec{E}\times\vec{B})\cdot\frac{1}{\|\phi\|^2}\varepsilon_{ab}\phi^a\nabla\phi^b
 \nonumber\\
 &&-\frac{1}{2m}(\alpha\vec{E}\times\vec{B})^2+\frac{\alpha}{2}E^2
 -\frac{g}{2}\|\phi\|^2-U-V,
\end{eqnarray}
where
\begin{equation}
 U=-\frac{\hbar^2}{2m}\frac{\nabla^2\|\phi\|}{\|\phi\|}
\end{equation}
is the Bohm quantum potential \cite{16}.

Differentiate Eq. (\ref{45}) with respect to coordinates. Using
(\ref{10}), it's easy to prove
\begin{eqnarray}
 -\frac{\hbar^2}{2m}\nabla\left(\nabla\frac{\phi^a}{\|\phi\|}\right)^2
 &=&m\vec{u}\times(\nabla\times\vec{u})
 -\nabla(\frac{1}{2}mu^2)
 \nonumber\\&&
 +\nabla[(\alpha\vec{E}\times\vec{B})\cdot\vec{u}]
 +\frac{1}{2m}\nabla(\alpha\vec{E}\times\vec{B})^2,
 \\
 \hbar\nabla\left(\frac{1}{\|\phi\|^2}\varepsilon_{ab}\phi^a\frac{\partial}{\partial
 t}\phi^b\right)
 &=&m\frac{\partial\vec{u}}{\partial t}
 -m\frac{\partial\vec{x}}{\partial t}\times(\nabla\times\vec{u}),
\end{eqnarray}
then we get the reduced dynamical equation
\begin{equation}
 m\frac{\partial\vec{u}}{\partial t}
 =m(\vec{u}+\frac{\partial\vec{x}}{\partial t})\times(\nabla\times\vec{u})
 -\nabla(\frac{1}{2}mu^2-\frac{\alpha}{2}E^2+\frac{g}{2}\|\phi\|^2+U+V).
\end{equation}

For stable vortex lines, $\frac{\partial\vec{u}}{\partial t}=0$, we
have
\begin{equation}
 m(\vec{u}+\frac{\partial\vec{x}}{\partial t})\times(\nabla\times\vec{u})
 =\nabla(\frac{1}{2}mu^2-\frac{\alpha}{2}E^2+\frac{g}{2}\|\phi\|^2+U+V),
\end{equation}
then
\begin{equation}\label{51}
 \nabla(\frac{1}{2}mu^2-\frac{\alpha}{2}E^2+\frac{g}{2}\|\phi\|^2+U+V)\cdot
 (\nabla\times\vec{u})=0.
\end{equation}

Just like the situation in fluid mechanics, on those vortex lines
\begin{equation}
(\nabla\times\vec{u})^i\propto dx^i.
\end{equation}
Eq. (\ref{51}) becomes
\begin{equation}
 \partial_i(\frac{1}{2}mu^2-\frac{\alpha}{2}E^2+\frac{g}{2}\|\phi\|^2+U+V)dx^i=0,
\end{equation}
thus
\begin{equation}
 \frac{1}{2}mu^2-\frac{\alpha}{2}E^2+\frac{g}{2}\|\phi\|^2+U+V=const.
\end{equation}
It's a conserved dynamics quantity on the stable vortex lines.

\section{Conclusions}\label{Sec5}

The use of $\phi$-mapping topological theory has clarified the issue
on the generation of vortices in quantum R\"ontgen effect. The
explicit expression of curl $\vec{u}$ is given, from which we get
the vorticity of the velocity field and the magnetic flux of the
monopoles, both of them are quantized by the product of the Hopf
indices and the Brouwer degrees. The generating, annihilating,
splitting and merging of vortex lines are discussed in detail, and
the speeds of these processes are estimated, which could come in
handy when experiments are concerned. The reduced dynamic equation
of the condensate is derived, from which we get a conserved dynamic
quantity on the stable vortex lines.

\section*{Acknowledgment}

This work was supported by the National Natural Science Foundation
of the People's Republic of China and the Fundamental Research Fund
for Physics and Mathematics of Lanzhou University.

\end{document}